# Highly-efficient fiber to Si-waveguide free-form coupler for foundry-scale silicon photonics

LUIGI RANNO,[1] JIA XU BRIAN SIA,[1,2] COSMIN POPESCU,[1] DREW WENINGER,[1] SAMUEL SERNA,[1,3] SHAOLIANG YU,[4] LIONEL C. KIMERLING,[1] ANURADHA AGARWAL,[5] TIAN GU,[1,5] AND JUEJUN HU[1,5,*]

[1]*Department of Materials Science & Engineering, Massachusetts Institute of Technology, 77 Massachusetts Avenue, Cambridge, MA 02139, USA*
[2] *Centre for Micro- & Nano-Electronics (CMNE), Nanyang Technological University, 50 Nanyang Avenue, Singapore 639798, Singapore*
[3] *Department of Physics, Photonics and Optical Engineering, Bridgewater State University, 131 Summer Street, Bridgewater, MA 02324, USA*
[4] *Zhejiang Laboratory, Hangzhou 311100, China*
[5] *Materials Research Laboratory, Massachusetts Institute of Technology, 77 Massachusetts Avenue, Cambridge, MA 02139, USA*
*\*hujuejun@mit.edu*

**Abstract:** As silicon photonics transitions from research to commercial deployment, packaging solutions that efficiently couple light into highly-compact and functional sub-micron silicon waveguides are imperative but remain challenging. The 220 nm silicon-on-insulator (SOI) platform, poised to enable large-scale integration, is the most widely adopted by foundries, resulting in established fabrication processes and extensive photonic component libraries. The development of a highly-efficient, scalable and broadband coupling scheme for this platform is therefore of paramount importance. Leveraging two-photon polymerization (TPP) and a deterministic free-form micro-optics design methodology based on the Fermat's principle, this work demonstrates an ultra-efficient and broadband 3-D coupler interface between standard SMF-28 single-mode fibers and silicon waveguides on the 220 nm SOI platform. The coupler achieves a low coupling loss of 0.8 dB for fundamental TE mode, along with 1-dB bandwidth exceeding 180 nm. The broadband operation enables diverse bandwidth-driven applications ranging from communications to spectroscopy. Furthermore, the 3-D free-form coupler also enables large tolerance to fiber misalignments and manufacturing variability, thereby relaxing packaging requirements towards cost reduction capitalizing on standard electronic packaging process flows.

## 1. Introduction

Driven by inherent advantages in highly scalable manufacturing [1,2], silicon photonics is a disruptive technology that has influenced multiple application spheres such as quantum optics [3], neuromorphic computing [4,5], optical communications [6], spectroscopy and sensing [7]. By virtue of the high-index contrast between silicon and its $SiO_2$ cladding ($\Delta n \sim 2$), dense integration have been enabled by the platform [8]. Active functionalities have also been implemented taking advantage of the high thermo-optic coefficient of silicon [9], its high-speed carrier modulation ability [2,10,11], and monolithic integration with fast and efficient germanium-based photodetectors [12]. In addition, leveraging the established CMOS fabrication processes already developed for the electronics industry, silicon stands as the unrivaled champion in the photonics space.

In recent years, we have witnessed growing commercial deployment of silicon photonics, owing to the increasing sophistication and functionality of silicon photonics circuits as well as critical advantages in manufacturing scalability [13–16]. However, photonic packaging is progressively becoming a major bottleneck. Custom tools with low throughput and active alignment practices are often demanded for photonic packaging [17]. As a consequence, the process accounts for up to 80% of the total costs of photonic modules. The high cost of photonic packaging is mostly attributed to the difficulty of coupling light into and out of the photonic circuit. That comes as a result of the large modal field mismatch between high-index-contrast waveguides and optical fibers, which presents a major barrier for implementation of high-throughput, low-loss coupling schemes [18,19]. Conventional methods involving either edge or grating coupling have their respective limitations [18,19]. For instance, edge couplers (i) can only be placed at the rather limited chip shoreline, preventing full utilization of the chip area and imposing severe design constraints (e.g., on positioning of contact pads); (ii) are difficult to test at the wafer scale, although some solutions have been proposed [20,21], and (iii) generally require sub-micron alignment tolerances that are not compatible with passive alignment approaches [19,22]. Grating couplers, on the other hand, are generally not as efficient and show strong wavelength dependence [19,23] due to the inherent wavelength sensitivity of optical diffraction. From a broader perspective, the prevailing I/O issue could potentially infringe upon the competitive advantage that photonic systems, since high coupling losses can escalate the photonic link power budget and overall energy consumption of the entire system. As a result, solving the photonic packaging bottleneck remains one of the most pressing challenges in the field. Overcoming the aforementioned limitations will enable significant cost reduction through leveraging the prevailing electronic packaging infrastructure.

Although the photonics community has been able to overcome some of the above-mentioned downsides pertaining to the conventional optical coupling approaches – for instance Tummidi and Webster [24] reported on the use of multiple layers of SiN to form an expanded edge coupler mode capable of achieving coupling loss as low as -0.4 dB when coupling to a standard SMF-28 fiber, while Ding *et al.* [25] demonstrated the use of metal back-reflectors below the Si device layer to produce fully-etched, apodized grating couplers with a coupling loss of ~0.6 dB and 3-dB bandwidth of 70 nm. These solutions nonetheless require processing steps and unconventional material stacks which are not currently available at photonic foundries [1,2,26]. Typically, photonic foundries offer edge couplers with insertion losses on the order of 1.5 dB when coupled with a lensed fiber with a mode field diameter of 3 µm, and grating couplers with coupling losses of approximately -3 dB and 1-dB bandwidth in the order of 30 nm [1,2], still a far cry from the state-of-the-art. Coupling approaches that deviate minimally from the conventional foundry material stack are essential to ensure facile and rapid technology upscaling.

Recently, Two Photon Polymerization (TPP) has been proposed as a potential approach to provide efficient coupling strategies for integrated photonics [27–29]. Unlocking the third geometric dimension as an additional degree of freedom allows TPP to uniquely produce intricate 3-D designs with sub-micron alignment accuracy, while being compatible with backend integration after the wafer fabrication has been completed. In addition, TPP-written structures have conferred optomechanical alignment benefits during photonic assembly [30–33]. Pioneering work has validated TPP as a viable technique with increasing commercial deployment for robust and versatile photonic packaging [34], having demonstrated optical coupling of waveguides to other waveguides [35–37], lasers [38], free-space [39,40], and fibers [27]. As an example, Lindenmann achieved coupling losses of approximately -4 dB using photonic wire bonding to link a waveguide to a single mode fiber [41] and -1.7 dB to a multicore fiber [42]. Luo and co-authors reported structures capable of achieving less than 1 dB insertion loss when coupling to single mode or multi-core fibers [43,44]. One limitation of these demonstrations is that they all involve a mode transformer transitioning from an uncladded on-chip waveguide to a low-index-contrast polymer waveguide section prior to coupling to the fiber to mitigate the mode size mismatch. The need for an exposed waveguide section without top cladding nonetheless entails an additional cladding strip process which is often non-standard in foundry manufacturing, and elevates risks of contamination and performance degradation.

Harnessing the TPP technique [28], this work proposes a solution to the photonic packaging bottleneck. Our prior work has involved the demonstration of free-form couplers for SiN waveguides [45]. This work aims to significantly advance the concept by demonstrating efficient coupling from fibers to tightly-confined, high-index-contrast Si waveguides using the free-form couplers. Si waveguide coupling incurs a significant technical barrier given the considerably larger mode size mismatch between the Si waveguide and a standard flat-cut (i.e., non-lensed) fiber, which not only compromises the coupling efficiency but also makes broadband and polarization-agnostic operation challenging.

This work addresses several salient aspects with regards to the scalable chip-to-fiber coupling of light for Si waveguides: 1) a highly-efficient free-form coupler for foundry-manufactured Si waveguides has been demonstrated ( 35 µm × 70 µm) using free-form micro-optics enabled by TPP, with coupling losses as low as 0.8 dB. We note that the size of the reflector section can be further reduced (~35 × 30 µm$^2$) by trimming it to the divergence angles from the Si facet (Fig. 1a) without incurring in additional losses. While the coupler is designed for the fundamental TE mode, efficient fundamental TM mode coupling is also shown to be possible where 1-dB bandwidth of 180 nm is demonstrated for TE and 1-dB bandwidth of 130 nm is facilitated for TM. The coupler operates based on optical reflection, designed via the Fermat's principle. While diffraction and refraction exhibit inherent wavelength-dependency, optical reflections constitute a chromatically agnostic phenomenon that induces ultra-low optical loss under the total internal reflection (TIR) regime. 2) The Si waveguide to fiber free-form coupler is compatible with cladded waveguides, enabling foundry compatibility. To the best of our knowledge, this aspect has not been addressed thus far amongst free-form couplers as evident by the abovementioned examples. 3) Different from grating couplers where the fiber is coupled at an angle (i.e., 9°), the fiber is surface-normal in this work. With the growing integration of silicon photonic circuits, there will be an increasing challenge in accommodating the high densities of electrical and optical inputs/outputs (I/O) in a single-plane. This situation will be alleviated when the fiber is perfectly vertically coupled to the Si waveguide to fiber free-form coupler. Furthermore, the ability for perfectly vertical fiber coupling implies compatibility with standard alignment tools used in electronic packaging. 4.) The free-form coupler is highly-resilient to fiber misalignments. This significantly relaxes the precision required for photonic packaging, enabling the application of standard pick-and-place tools and processes from electronic packaging. In addition, the coupler is also tolerant to manufacturing variabilities. 5.) The reflector indicates good structural resilience to temperatures as high as 250 °C, indicating compatibility with solder reflow [17,46]. 6.) The broad bandwidth of the coupler enables far-reaching wavelength diverse applications such as neuromorphic computing [4,5], wavelength division multiplexing (WDM) communications [47], or spectroscopic sensing [48]. Lastly, we propose an approach that enables the scalable packaging of the reflectors for resilient fiber or fiber array bonding, of relevance for the real-world deployment of this technology.

## 2. Device concept and design

### 2.1 Broadband reflector design via the Fermat's principle

The Si waveguide to fiber 3-D free-form coupler, designed for the fundamental TE mode, is illustrated in Fig. 1(a). The Si taper, deep trench and reflector sections are indicated. The inverse taper was implemented by virtue of its relatively large output mode size (~3 µm$^2$), and hence, lower divergence angle as compared to a conventional waveguide. This result in a reduction in the aspect ratio of the reflector while simultaneously facilitating efficient total internal reflection by expanding the reflector size along the light propagation direction. The 3-D free-form coupler (Fig. 1(a), xyz coordinates labeled) utilizes the principle of wavefront interference to efficiently couple light between the in-plane Si waveguide mode and an out-of-plane fiber mode oriented orthogonally. To avoid computationally intensive point-by-point local optimization often used for the development of free-form optics [49], we take advantage of the Fermat's principle to vastly simplify the optimization procedure [45]. Considering an arbitrary reflector surface which redirects the lightwave exiting a waveguide upward into an optical fiber; the Fermat's principle dictates that the total optical path length traversed by the lightwave must be stationary, resulting in the following relation:

$$\phi_1 + \phi_2 = \phi_{total} = \text{constant} \quad (1)$$

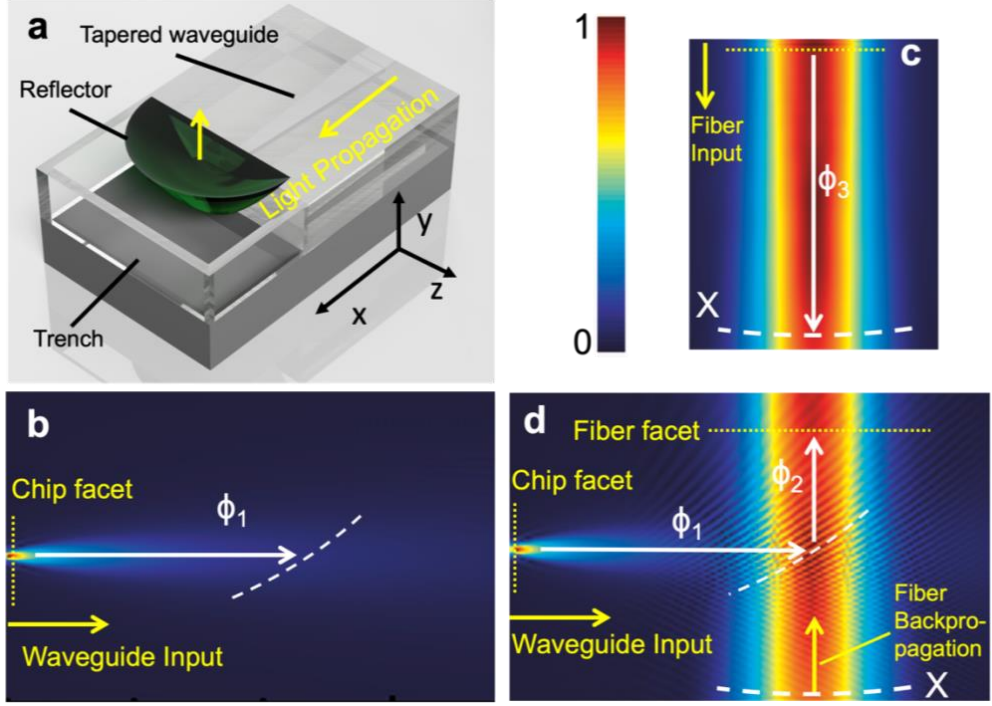

Fig. 1. (a) 3-D schematic of the fiber to Si waveguide free-form coupler. The Si nonlinear inverse taper, deep trench and reflector sections are annotated. The design methodology of the free form coupler is guided by the Fermat's principle. 3-D FDTD simulations of (b) forward propagating lightwave exiting from the Si waveguide inverse taper, (c) forward propagating lightwave from the fiber, and (d) co-propagation of back-propagated lightwave from the fiber and forward propagating light from the waveguide taper. The white arrows illustrate the phase accumulated by the lightwave as it propagates. The yellow arrows indicate the light sources in the different simulations, and corresponding propagation directions.

$\phi_1$ and $\phi_2$ are the phase delays accumulated by propagating the lightwave as it travels from the Si taper to the free-form reflector surface, and from the reflector surface to the fiber facet, respectively. $\phi_{total}$ must be a constant following the Fermat's principle. In Eq. 1, we neglect the TIR phase delay accumulated upon reflection at the coupler surface since it remains nearly constant for different light trajectories.

Figures 1(b)-(c) depict the phase of the propagating lightwave, determined through 3-D finite-difference time-domain (FDTD) modeling of the Si taper and optical fiber separately. In Fig. 1(d), we consider co-propagation of forward-propagating light from the Si taper and backward-propagation of the lightwave output from the fiber (i.e., the time reversal of Fig. 1(c)). To back propagate the lightwave from the fiber, we instead define a wavefront X (Fig. 1(c)) acting as the "light source" from which the lightwave propagates upward. $\phi_3$ denotes the phase delay from X to the fiber. According to Fig. 1(c), the phase delay due to propagation from the X to a point on the reflector surface is then given by $\phi_3 - \phi_2$. The condition for constructive interference between the lightwaves exiting from the waveguide and from the "source" X is:

$$\phi_1 - (\phi_3 - \phi_2) = \phi_{total} - \phi_3 = 2N\pi \quad (2)$$

Since $\phi_3$ is identical for all light paths between X and the fiber facet, Eq. 2 implies that $\phi_{total} = 2N\pi + \phi_3$ is a constant and is equivalent to the stationary optical path condition specified by the Fermat's principle. Therefore, the loci of constructive interference, which are located using a spiral trace search algorithm [45], correspond to a group of reflector surfaces that fulfill Eq. 1. The optimal surface can be identified by finding the locus yielding highest overlap integral between the two lightwaves (from the waveguide and X, respectively). Unlike prevailing grating couplers that involve computationally intensive optimization algorithms [26,50], this design approach involves only two full vectorial FDTD simulations, namely forward propagation of the fiber mode in absence of the Si waveguide, and co-propagation of the back-propagated (fiber) wave from X and light exiting from the Si waveguide mode. The optimal reflector surface is selected deterministically without iterative optimization.

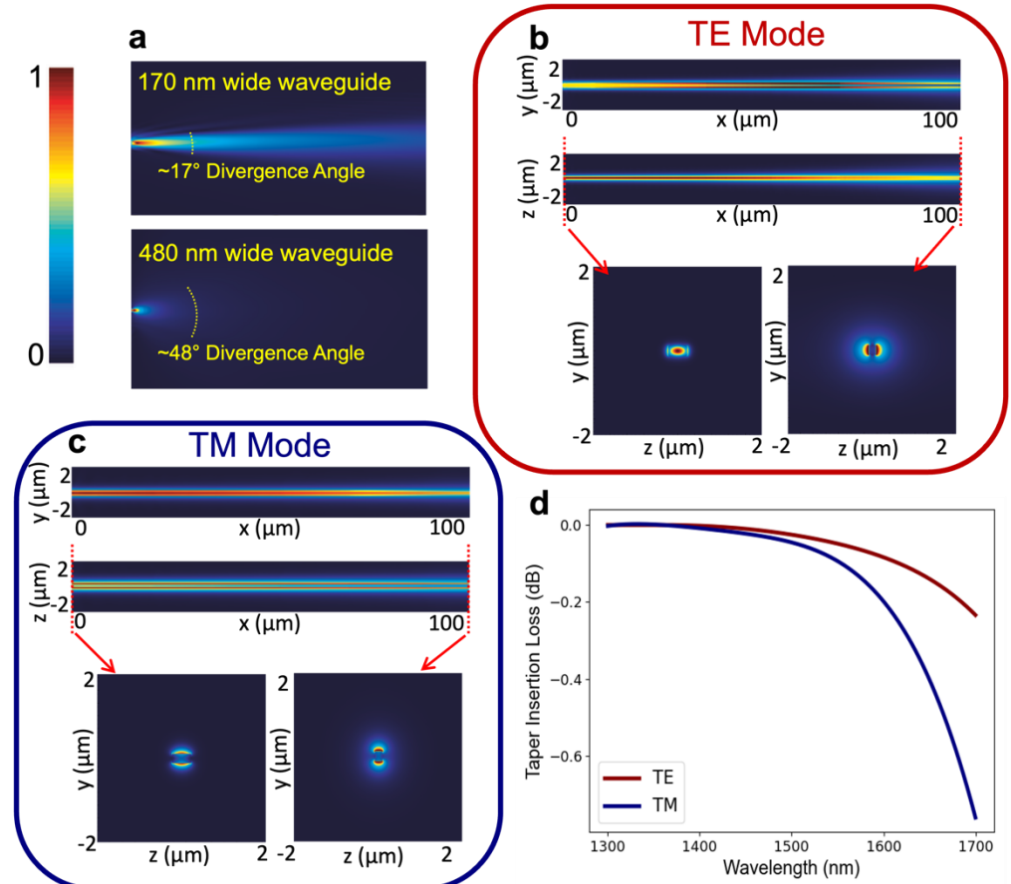

Fig. 2. (a) TE-mode electric field distribution in the xy-plane of lightwaves exiting from 170 and 480 nm wide waveguides at 1550 nm wavelength. The full-width at half-maximum divergence angles are marked. The electric field intensity of the lightwaves in the x-y and x-z planes as (b) fundamental TE, and (c) fundamental TM mode propagate along the nonlinear inverse taper. The respective electric field distributions (y-z plane) at both facets of the taper are shown. (d) Spectral dependence of the total insertion loss of the taper.

The Si waveguide geometry at the taper end facet plays a significant role in determining the shape and efficiency of the coupler (Fig. 1(a)). Figure 2(a) plots the electric field distribution of the lightwaves in the xy-plane as it exits from 170 (top) and 480 nm (bottom) wide waveguides. Here the 480 nm width is chosen because it is close to single-mode cutoff at the C-band. The narrower waveguide gives rise to a much lower divergence angle, as illustrated in Fig. 2(a). To that effect, reduction of the divergence angle through appropriate waveguide taper design will minimize loss due to asymmetric intensity distribution of the field reflected from the coupler [51]. It also follows that the minimum size of the reflector is bound by the highest angle at which TIR can be maintained, and given that the refractive index of typical TPP polymers is less than 1.6 [52], shallower structures enabled by tapered waveguides perform better.

The Si nonlinear inverse taper was designed to adiabatically expand the waveguide mode, thereby simultaneously reducing the coupler size and also the far-field divergence angle. Furthermore, the Si nonlinear inverse taper design critically impact the substrate leakage loss, as the Si waveguides are fabricated on a 220 nm SOI platform with 2 μm-thick buried oxide. By imposing a slower width tapering rate at the beginning and faster width change towards the end, the waveguide mode is confined as much as possible along the taper and only expanded at the end to facilitate coupling to the reflector, thereby enabling adiabatic transition with minimal substrate leakage loss. More specifically, the taper narrowing follows an $x^{1.25}$ function in between the two ends, where the exponent was optimized to minimize the total insertion loss of the taper. The length of the taper is 100 μm with an optimized insertion loss of ~0.05 dB at 1550 nm. The electric field distributions of the waveguide mode in the xy and xz planes are plotted. The nonlinear taper preserves the polarization while expanding the field laterally, hence resulting in a more symmetric mode profile. While the free-form coupler is designed for efficient coupling of the fundamental TE mode, it is anticipated that it can also operate with the fundamental TM mode. As such, the capability of the nonlinear inverse taper for adiabatic TM mode conversion is also investigated in Fig. 2(c), where low-loss mode conversion and mode spatial preservation is verified. The spectral dependences of the nonlinear inverse taper for both polarizations are considered in Fig. 2(d), showing that the substrate leakage losses dominate at longer wavelengths. The fundamental TM mode shows larger substrate leakage loss as compared to TE, due to the stronger out-of-plane confinement of the TE mode field.

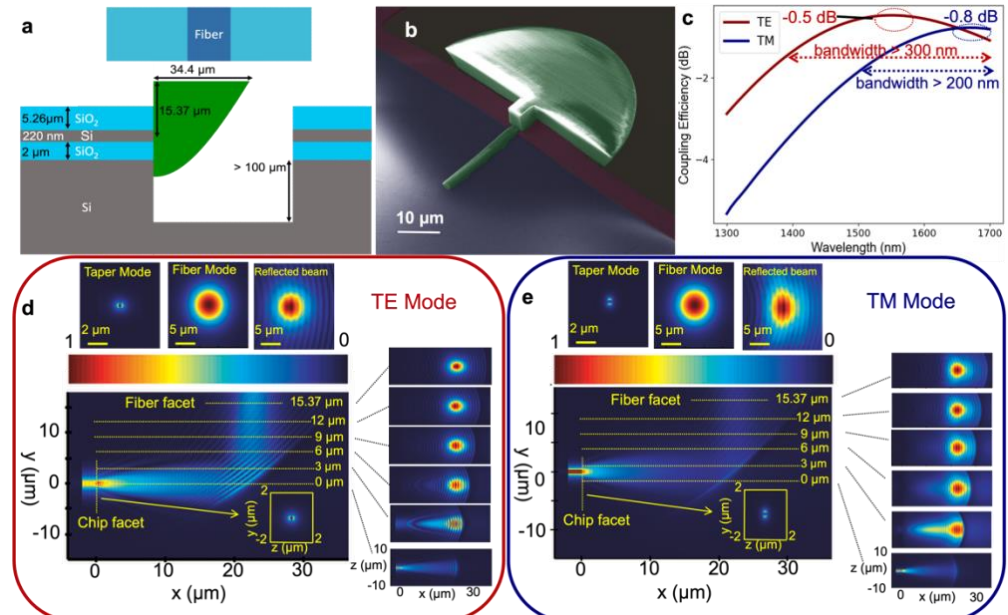

Fig. 3 (a) Cross-sectional illustration of the designed free-form coupler, where the constituent components are illustrated, and critical dimensions labelled. (b) False-colored SEM image of the fabricated free-form coupler. The two slabs (short and thick, and long and thin) extending towards the waveguide are visual aids to quickly gauge alignment accuracy in and out-of plane, they do not perform any optical function. (c) Simulated spectral performance of the designed free-form coupler for the case of TE and TM waveguide polarizations. Electric-field intensity (xy-plane) as the lightwave is coupled between the fiber and the Si waveguide through the coupler for the case of (d) fundamental TE, and (e) TM polarization. The electric field (xz-plane) evolution of the optical mode as it is reflected normal from the Si waveguide and propagates towards the fiber facet is shown for different y positions. The optical modes at the taper, reflector and fiber output are also illustrated.

The xy-section of the free-form coupler is shown in Fig. 3(a). The photonic chip was fabricated at AIM Photonics, whose standard process design kit (PDK) includes etching a deep trench for dicing. Here we leverage the etching step to define the recess in which the reflectors were printed. A false colored SEM image is shown in Fig. 3(b). The reflectors were fabricated using TPP by polymerizing IP-n162, a commercially available resin from Nanoscribe GmbH boasting low absorption and high TPP resolution. The printing process was carried out using a Photonic Professional GT2 system with an averaged laser power of 60 mW and hatching/slicing distances of 100 and 200 nm, respectively. Top panels in Fig. 3(d) show the near-field optical intensity distributions for TE polarization at the waveguide end facet, the fiber mode and the reflected beam. A large field overlap of 92% between the latter two was verified through overlap integral. The beam forming characteristics of the reflector are showcased by plotting xz-sections of optical intensity distributions at different heights from the reflector surface. For comparison, the near-field distributions with TM-mode input are also shown in Fig. 3(e). For TM polarization, a field overlap of 84% between the reflected beam and the fiber mode is achieved. The spectral performance of the coupler is modeled and plotted in Fig. 3(c) for the fundamental TE mode, where coupling loss as low as 0.5 dB is shown. While the free-form coupler is designed and optimized for the fundamental TE mode, it is found to be also efficient for the fundamental TM mode where minimum coupling loss as low as 0.8 dB (1.3 dB at 1550 nm) is computed. Through simulation, the coupler shows 1-dB bandwidth of over 300 nm in the TE polarization, and 1-dB bandwidth exceeding 200 nm in the TM polarization. The electric field distributions in the xy-plane are shown in Fig. 3(d) and (e) respectively for the two polarizations. Even though the coupler is designed for the TE mode it still affords highly-efficient coupling for both polarizations compared to the state-of-the-art [1,2,19].

*2.2 Fiber-coupler alignment tolerance analysis*

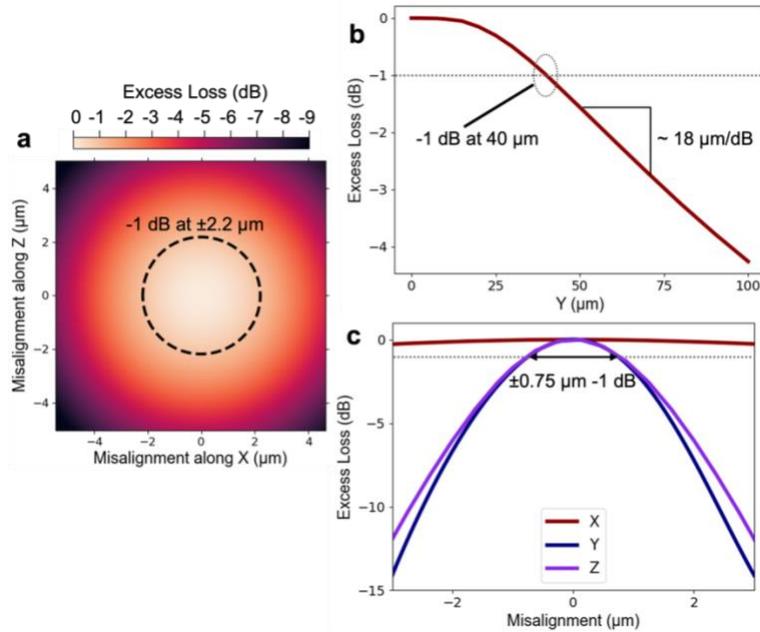

Fig. 4. Computed fiber misalignment tolerance in in the (a) xz-plane, and (b) y-direction. (c) Computed excess loss in coupling efficiency of the reflector as a result of printing misalignments with respect to the waveguide facet. Simulated for TE polarization at 1550 nm.

Tolerance to fiber misalignment is an important metric with regard to photonic packaging, as a more tolerant coupler allows using faster packaging procedures that forego the burden of active alignment and drive down the packaging cost. In Figs. 4(a)-(b), the alignment tolerance of the 3-D free-form coupler is investigated through simulations. From Fig. 4(a) it can be seen that the 1-dB alignment tolerance of the coupler with respect to in-plane (x-z) fiber misalignments is approximately ±2.2 μm, as expected given the beam size closely matching the SMF-28 fiber mode. In addition, the coupler shows remarkable tolerances in the y-direction, where a spacing as large as 40 μm between the fiber and the coupler leads to an excess loss of only 1 dB, much better than the 10 μm that conventional grating couplers achieve [53]. This can be attributed to the highly collimated beam exiting from the top surface of the coupler. The tolerance to misalignment between the coupler and the waveguide, which can occur during the TPP printing process, is also investigated in Fig. 4(c). The design is highly tolerant to misalignment in the x-direction whereas ±6.5 μm alignment accuracy is mandated for an excess loss of less than 1 dB.

Furthermore, the coupler also shows good resilience to manufacturing tolerance in the y and z-direction, where misalignment of ±0.75 μm leads to an increase in excess loss of 1 dB. These values are well within the alignment accuracy of TPP: printing with misalignment below 200 nm have been experimentally demonstrated [54] and commercial systems capable of automated alignment are already available on the market [55]. The results suggest excellent robustness of the free-form coupler technology against possible alignment errors in both manufacturing and packaging processes.

## 3. Experimental data

The transmission of the 3-D free form coupler is measured using two flat-cleaved fibers (SMF-28), oriented normal to the chip surface, directed to two couplers located at both ends of a silicon waveguide (Fig. 5(a)), and controlled through the use of piezoelectric motors with an automated stage (SD-100 from Mapleleaf Photonics). As a testament to their wavelength-agnostic performance, Fig. 5(a) demonstrates coupling of visible red light with less than 1 dBm input power to the waveguide that, despite the significant material losses of silicon, a bright beam output beam is visible. Three different tunable laser sources (TSL series from Santec) are controlled through the use of an optical switch module (OSU-100), which is connected to the input fiber via a fiber polarization controller (FPC), while the photodetector (MPM-210) is connected to the end of the output fiber. The measured spectral performance of an individual reflector is shown in Fig. 5(b), where the coupling loss is taken as half the total measured insertion loss after subtraction of waveguide propagation loss and system loss. A minimum coupling loss of 0.8 dB with a 1-dB bandwidth in excess of 180 nm was observed for the TE mode. The TM polarization yields a minimum loss of ~3 dB per facet, with a 1-dB bandwidth larger than 130 nm, despite the fact that the coupler is designed and optimized for the TE polarization. The experimental results nonetheless still indicate excellent dual polarization coupling capabilities, especially considering that typically TM grating couplers have insertion losses of ~ 4-5 dB [56,57]. The experimental results correspond closely to the designed performance, which validates the coupler fabrication process flow. To the best of our knowledge the free-form coupler demonstrates one of the best fiber to Si waveguide coupling performance (coupling loss and bandwidth) based on a foundry-compatible design.

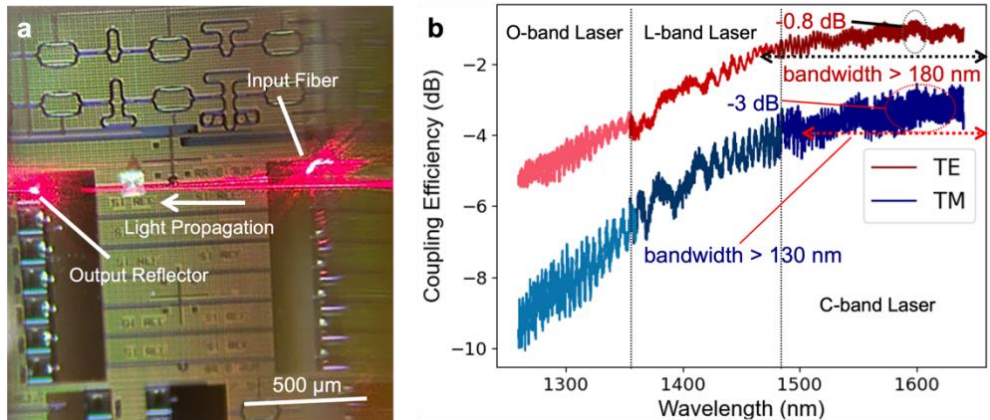

Fig. 5. (a) Micrograph showing fiber-coupled (SMF-28) red laser light (630 nm) into a Si waveguide via the free-form couplers. (b) Measured spectral performance of the 3-D free-form coupler pertaining to the fundamental TE and TM modes.

Following from the simulations performed in Fig. 4, the fiber misalignment tolerance was measured (Fig. 6) using a piezoelectric actuator with a step size of 500 nm. After determining the optimal position of the input and output fibers, the position of the output fiber was rastered to map the coupling efficiency as a function of the fiber position. The coupler is experimentally verified to enable 1-dB alignment tolerances of ~ ±2 μm in the xz-plane (Fig. 6(a)), consistent with simulation results. Furthermore, as expected from the collimated beam formed by the coupler, a large y-direction (Fig. 6(b)) tolerance is also observed, where a 19 μm misalignment only leads to 1 dB excess loss.

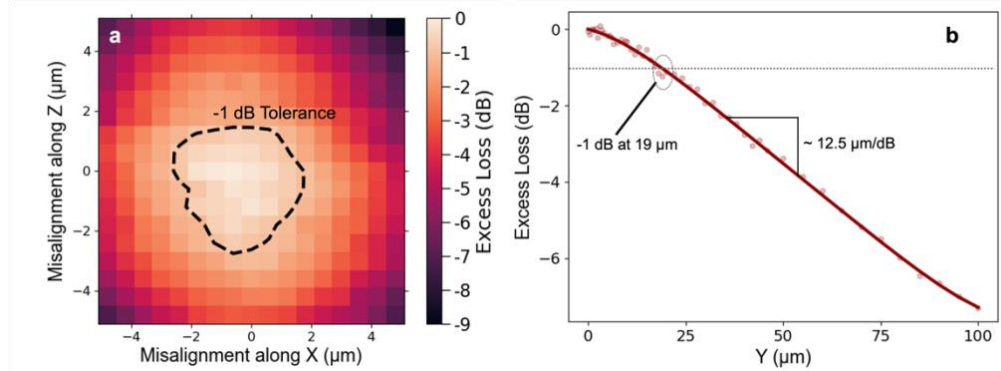

Fig. 6. (a) Measured fiber misalignment tolerance in the xz-plane measured using a piezoelectric actuator with a step size of 500 nm. The contour traces the 1-dB alignment tolerance. (b) Measured excess loss as a function of output fiber misalignment in the y-direction.

## 4. Discussion

To demonstrate the advantages of the proposed design, the free-form coupler is compared against a standard foundry-PDK grating coupler (Fig. 7(a)). The transmission across a Si waveguide configured with grating couplers on both ends, as well as one with a grating and a free-form coupler, were measured using the LUNA OVA 5000 optical vector analyzer and shown in Fig. 7(a). The insertion loss of the setup and waveguide propagation loss were subtracted from the two spectra. The advantages of the 3-D free form coupler can be clearly expounded. First, ~2 dB decrease in insertion loss is evident even at the peak efficiency of the gratings. Furthermore, significant broadening of operation bandwidth can be seen. The increase in bandwidth becomes even more apparent when compared to Fig. 5(b), where the couplers are implemented on both ends of a waveguide. This speaks of the superior fiber to chip coupling capabilities of the free-form couplers for potential deployment in next-generation silicon photonic circuits where power consumption is of upmost importance.

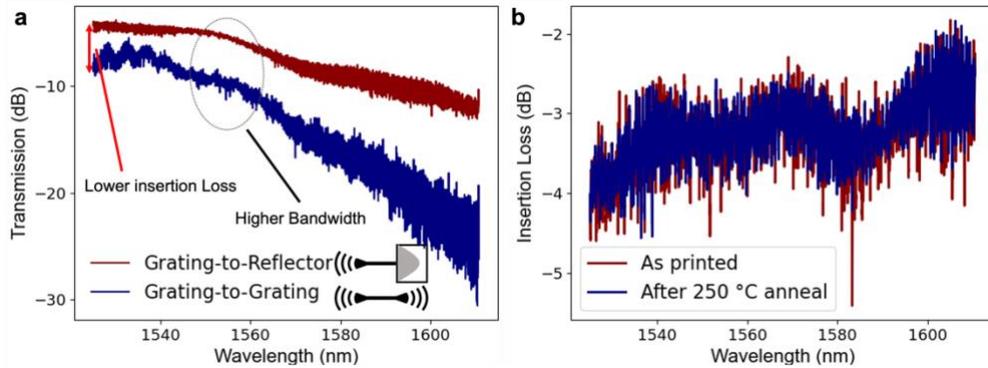

Fig. 7. (a) Measured spectra for two Si waveguides with different configurations of light coupling: grating-to-reflector and grating-to-grating. (b) Measured total insertion loss of a Si waveguide terminated with two free-form couplers as-fabricated and after annealing at 250 °C for 10 minutes.

In addition to their excellent optical performances, the free-form couplers are also compatible with standard electronic packaging processes. This is in view of 1) the surface-normal (i.e., with zero tilt angle) fiber coupling configuration that is compatible with standard electronic packaging tools; 2) the significant tolerance to fiber misalignment (1-dB alignment tolerances of ±2 µm and 19 µm in the xz plane and y-direction, respectively), commensurate with state-of-the-art assembly tools used in electronic packaging [58,59]; and 3) compatibility with the thermal budget of standard packaging processes. The last point was validated by comparing the measured spectra of the couplers before and after annealing (250 ˚C) for 10 minutes in Fig. 7(b). The conditions were chosen to represent the typical heat treatment process used in Pb-free solder reflow [46].

In addition to the benefits presented above, the free-form couplers feature a compact footprint and are designed to work with standard foundry-processed cladded waveguides (i.e., do not require a long uncladded or polymer cladded section for mode transition [42–44,60]). This is an important advantage over other proposed TPP couplers enabling high-density optical I/O on 'zero-change' foundry-manufactured photonic chips. The coupler can also be combined with TPP-printed optomechanical fixtures to facilitate passive alignment and assembly. Some excellent work done by Jimenez Gordillo *et al.*[30] showcased the potential of TPP for such purposes, although only it was only applied to grating couplers. Similarly, the Bakir research group has investigated various approaches to interconnect fibers and photonic chips (with grating couplers) using etched chips with printed fiber ferrules [31–33]. We propose a potential solution for large-scale packaging of the free-form couplers with fiber arrays. The envisioned structure is illustrated in Fig. 8, both for the case of an individual optical fiber (Fig. 8(a)), and a fiber array (Fig. 8(b)). After printing of the reflector(s), an additional alignment structure is printed onto the chip top surface, maintaining alignment with the printed reflectors. The alignment structures consist of two components: one guiding sleeve which serves the purpose of guiding the fiber (array) into the position, and a stopper which prevents collision with the reflector top surface and also enables optimal spacing between the fiber facet and reflector top surface. Once the fiber (array) is rested onto the stopper, optical epoxy can be used to bond the assembly together, forming a sturdy package capable of withstanding mechanical vibrations while also isolating the reflectors from the external environment. Printing the bulk of the alignment structure, which does not require high resolution, may take advantage of beam expansion procedures that vastly increase the voxel size (e.g., UpNano's adaptive resolution function [61]), hence substantially reducing the total printing time. In addition, the lack of overhanging features (both the reflectors and alignment structures) makes the proposed design stand out thanks to its compatibility with nanoimprint/embossing. This convenient and scalable packaging solution differentiates our design from prior arts, opening the door to future implementations at the wafer scale.

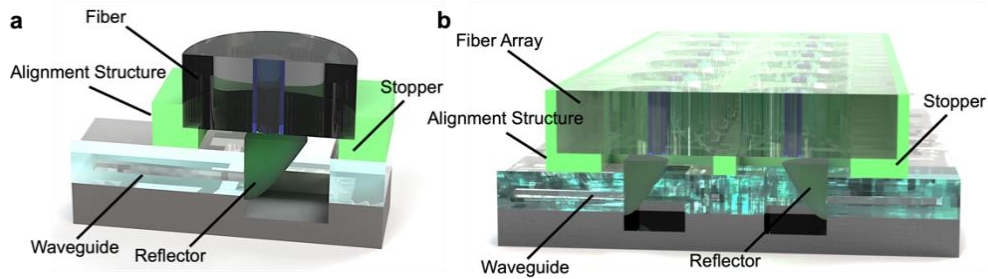

Fig. 8. Potential photonic packaging solution involving the freeform coupler, enabling packaging with (a) individual fibers and (b) fiber arrays. A mechanical alignment feature is printed onto the chip, with a stopper to prevent collision of the fiber (array) with the reflector surface. The fiber (array) is guided by the alignment structure and brought down onto the chip, until it rests onto the stopper. The optical assembly can then be bonded with optical epoxy as is conventionally done for grating couplers.

## 5.  Conclusion

This work demonstrated 3-D free-form couplers that enable highly-efficient and broadband coupling between fibers and tightly-confined Si waveguides. The 3-D free-form couplers, optimized for the TE mode, lead to a coupling loss down to 0.8 dB in the C-band with a 1-dB bandwidth in excess of 180 nm. In addition, the coupler can also operate with TM polarization with a minimum coupling loss of 3 dB and a 1-dB bandwidth of larger than 130 nm. The free-form couplers also feature significant tolerance against fiber and manufacturing misalignments, thereby relaxing the precision and cost involved in photonic packaging. The coupler design is fully compatible with the standard 220 nm SOI platform without involving custom cladding stripping steps, and is compatible with the thermal budget of solder reflow processes commonly used in electronic/photonic packaging. We envisage this work to provide a viable pathway to resolve the pressing issues related to the packaging bottleneck that silicon photonics is facing.

**Funding.** This work was partially supported by NSF ITE Convergence Accelerator Track I: Building a Sustainable, Innovative Ecosystem for Microchip Manufacturing, Award Number ITE-2236093. J.X.B. S. acknowledges financial support from the Ministry of Education - Singapore (International Postdoctoral Fellowship).

**Acknowledgments.** Part of this work was conducted at the shared nanofabrication facilities of MIT.nano and Harvard's Center for Nanoscale Systems (CNS).

**Disclosures.** The authors declare no conflicts of interest.

**Data availability.** Data underlying the results presented in this paper are not publicly available at this time but may be obtained from the authors upon reasonable request.